\documentclass{article}

\usepackage{microtype}
\usepackage{graphicx}
\usepackage{subfigure}
\usepackage{booktabs} 

\usepackage{hyperref}
\usepackage{cuted}
\usepackage{chemformula}
\usepackage{pifont}%
\usepackage{amsmath}
\usepackage{amssymb}
\usepackage{mathrsfs}
\newcommand{\angstrom}{\mbox{\normalfont\AA}}

\usepackage[accepted]{icml2023}

\usepackage{amsmath}
\usepackage{amssymb}
\usepackage{mathtools}
\usepackage{amsthm}

\usepackage[capitalize,noabbrev]{cleveref}

\theoremstyle{plain}

\theoremstyle{definition}

\theoremstyle{remark}

\usepackage[textsize=tiny]{todonotes}

\icmltitlerunning{Molecular Geometry-aware Transformer for accurate 3D Atomic System modeling}

\begin{document}

\twocolumn[
\icmltitle{Molecular Geometry-aware Transformer \\ for accurate 3D Atomic System modeling}




\begin{icmlauthorlist}
\icmlauthor{Zheng Yuan}{yyy}
\icmlauthor{Yaoyun Zhang}{yyy}
\icmlauthor{Chuanqi Tan}{yyy}
\icmlauthor{Wei Wang}{yyy}
\icmlauthor{Fei Huang}{yyy}
\icmlauthor{Songfang Huang}{yyy}
\end{icmlauthorlist}

\icmlaffiliation{yyy}{Alibaba Group}

\icmlcorrespondingauthor{Songfang Huang}{songfang.hsf@alibaba-inc.com}

\icmlkeywords{Molecular modeling, 3D Transformer}

\vskip 0.3in
]

\printAffiliationsAndNotice{}





\begin{abstract}
Molecular dynamic simulations are important in computational physics, chemistry, material, and biology.
Machine learning-based methods have shown strong abilities in predicting molecular energy and properties and are much faster than DFT calculations. 
Molecular energy is at least related to atoms, bonds, bond angles, torsion angles, and nonbonding atom pairs.
Previous Transformer models only use atoms as inputs which lack explicit modeling of the aforementioned factors.
To alleviate this limitation, we propose \textbf{Moleformer}, a novel Transformer architecture that takes nodes (atoms) and edges (bonds and nonbonding atom pairs) as inputs and models the interactions among them using rotational and translational invariant geometry-aware spatial encoding. 
Proposed spatial encoding calculates relative position information including distances and angles among nodes and edges.
We benchmark Moleformer on OC20 and QM9 datasets, and our model achieves state-of-the-art on the initial state to relaxed energy prediction of OC20 and is very competitive in QM9 on predicting quantum chemical properties compared to other Transformer and Graph Neural Network (GNN) methods which proves the effectiveness of the proposed geometry-aware spatial encoding in Moleformer.
\end{abstract}


\section{Introduction}
Molecular dynamic simulations (MDS) are essential in many subjects including physics, chemistry, biology, and materials \cite{karplus1990molecular,van1990computer,md,luding2005molecular,durrant2011molecular}.
Density function theory (DFT) is the most commonly used computational tool for simulating atomic systems and estimating the energy of molecular systems \cite{dft}.
While DFT can provide highly accurate estimations, it is extremely slow to apply to large atomic systems due to numerous iterate steps \cite{dft2}.
Machine learning methods can estimate molecular system energy from \textit{ab initio} states which do not need to inference multiple times and have satisfied performances.
Machine learning methods use positional information and atomic numbers of atoms and estimate potential energy using neural networks \cite{schnet,graphormer3d}. 

The basic form of force field for the molecular system energy can be decomposed into a sum of functions based on factors including bonds, bond angles, torsion angles, and nonbonding interactions (including electrostatic and van der Waals) \cite{gasteiger_dimenet_2020,cookbook}.
To include aforementioned factors for energy estimation, a neural network needs to learn interactions among atom-atom (forming bonds and nonbonding atom pairs), atom-bond (forming bond angles), and bond-bond (forming torsion angles). Previous Transformer-based models learn interactions among atom-atom in each Transformer layer \cite{graphormer3d,Equiformer}. Since one Transformer layer cannot model interactions of more than two atoms, high-order interactions (interactions among three or four atoms) are only modeled implicitly across Transformer layers.

To alleviate this limitation and leverage the power of Transformers, we propose Moleformer which explicitly models interactions among atoms and atom pairs in each Transformer layer. Moleformer takes atoms, bonds, and non-bonding atom pairs as inputs and applies a novel translational and rotational invariant geometry-aware spatial encoding to capture geometry relations among these inputs. The proposed spatial encoding is calculated by relative distances and angles of the atoms and atom pairs which is enough to determine the relative spatial geometry information among them. Interactions among atom-bond include the bond angle factor from the force field, and interactions among bond-bond include the torsion angle factor in the force field.  
DimeNet \cite{gasteiger_dimenet_2020} had introduced directional embeddings which learn bond representation from the neighbor bonds. Compared to DimeNet, Moleformer can further learn atom pairs representation from nonadjacent atoms and atom pairs which may better model the nonbonding interactions introduced from the force field.

We demonstrate our method on Open Catalyst 2020 (OC20) dataset \cite{oc20} and the QM9 dataset \cite{qm9} and achieve competitive results to previous methods.
Moleformer achieves state-of-the-art performance in predicting catalyst system relaxed energy from \textit{ab initio} states using only \textit{ab initio} states training data.
We also propose a periodic boundary condition correction missed in previous machine learning methods which can improve the performances of relaxed states and energy prediction in OC20. 
Moleformer also performs better than recent state-of-the-art methods including SEGNN \cite{segnn} and EQGAT \cite{eqgat} in QM9 quantum chemical properties predictions and shows strong abilities to predict molecular orbital energy.




\section{Related Work}
Machine learning-based MDS methods are gaining increasing attention due to their fast inference speed and good performance \cite{schnet,zhang2020molecular,gasteiger_dimenetpp_2020,painn,hmgnn,wang2022heterogeneous,doi:10.1021/acs.jcim.0c01224}.
Recent works in this direction can be mainly divided into two groups according to backbone neural network architectures: Graph Neural Networks (GNNs) \cite{gasteiger_dimenet_2020,hu2021forcenet,gemnet,spinconv,scn} and Transformers \cite{se3,graphormer,graphormer3d,Equiformer}.
GNNs construct graphs based on atom positions treating atoms as nodes and bonds as edges.
Node representations are iteratively updated by messages aggregated nodes neighborhoods.
However, GNNs can only interact with pre-defined neighborhoods in each layer.
Compared to GNN models, Transformers can model interactions between all atom-atom pairs through the self-attention mechanism in each Transformer layer \cite{vaswani2017attention}, which enables the model to learn complex relations between nonbonding atom pairs. 
Graphormer \cite{graphormer,graphormer3d} takes atoms as inputs and modifies absolute positional encoding in Transformers by using the relative distance between atoms to represent geometric information.
Different from Graphormer, Moleformer uses atoms and edges as inputs and uses the relative geometry information between atoms and edges.


\section{Methods}

Consider a 3D atom system with $N$ atoms (i.e. nodes) described by nuclear charges $\{z_1,...,z_N\}, z_i\in\mathbb{Z}$ and 3D positions $\{\mathbf{x}_1,...,\mathbf{x}_N\}, \mathbf{x}_i\in\mathbb{R}^3$.
The target is to predict scalar property $y\in\mathbb{R}$ of the system (e.g. system energy, dipole moment, and heat capacity).

\subsection{Architeture of Moleformer}

\begin{figure*}[ht]
    \centering
    \includegraphics[width=0.9\linewidth]{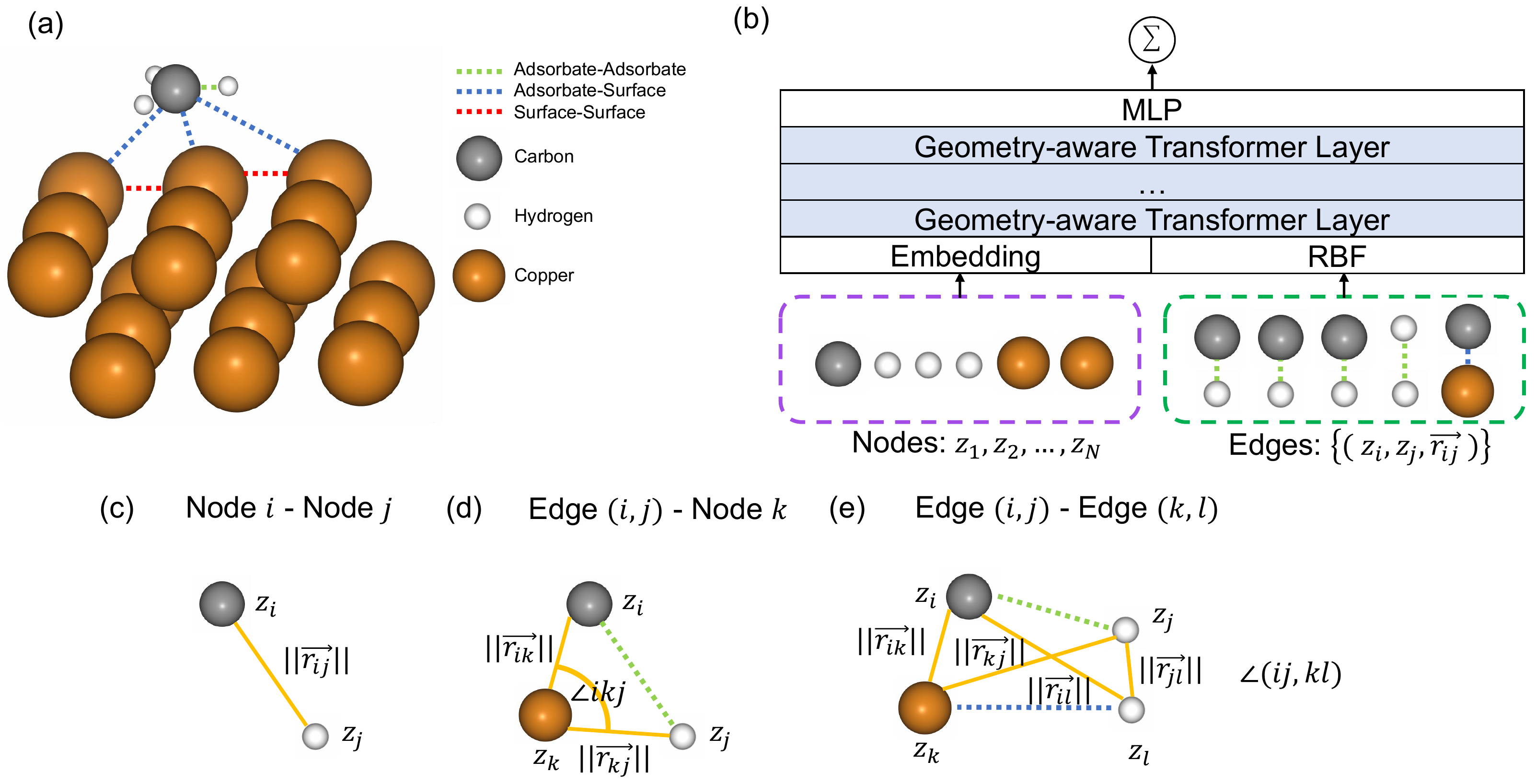}
    \caption{\textbf{The overview of Moleformer.} \textbf{(a)} A toy example of a catalyst system where \ch{CH3} is the adsorbate and the surface is composed of copper atoms. Moleformer prefers edges between adosrbate-adsorbate atoms and adsorbate-surface atoms as model inputs. \textbf{(b)} Architecture of Moleformer: Nodes and selected edges are embedded by an embedding layer and an RBF layer. Stacks of geometry-aware transformer layers are used for interacting information from nodes and edges. An MLP layer is used for generating node-level or edge-level outputs, and the scalar property is predicted by summation of them. \textbf{(c)-(e)} show how geometry-aware spatial encoding interacts among node-node, edge-node, and edge-edge in transformer layers respectively. Dotted lines denote edges and solid yellow lines denote geometry information (i.e. distances and angles) used in the spatial encoding.
    \textbf{(c)} Distances between node-node are used. \textbf{(d)} Distances and angles between edge-node are used. 
    \textbf{(e)} Distances and angles between edge-edge are used.
    }
    \label{fig:head}
\end{figure*}

Moleformer (Figure~\ref{fig:head} \textbf{(b)}) is a rotation- and translation-invariant Transformer \cite{vaswani2017attention} for 3D molecular energy and property prediction.
Moleformer uses nodes (atoms) and selected edges (including bonding and nonbonding atom pairs) as  inputs.
If an atomic system contains $N$ atoms, there are $C^2_N$ possible edges.
Involving all possible edges into the models is computational-heavy and may introduce uninformative edges.
We choose atom pairs as edges with top-$M$ closest distances.
For an atomic system like a catalyst system comprising an adsorbate and a surface (Figure~\ref{fig:head} \textbf{(a)}), we prefer edges between adsorbate-adsorbate atom pairs and adsorbate-surface atom pairs rather than surface-surface atom pairs.
The reason is that the edges between such atom pairs are more diverse in edge lengths and atom types and may contain more information in energy modeling.
We will penalize distances in surface-surface atom pairs for edge selection.


\paragraph{Input Embedding}

We embed edge $(i,j)$ by nuclear charges $z_i,z_j$ and distances between two atoms $\|\overrightarrow{r_{ij}}\|= \|\mathbf{x}_i - \mathbf{x}_j\|$ into $h_{ij}^0$ using RBF \cite{schnet}.

\begin{equation}
    h_{ij}^0 = {\rm RBF}(z_i,z_j,\|\overrightarrow{r_{ij}}\|)
\end{equation}

Following \citet{graphormer3d}, we embed atom $i$ into $h_i^0$ by nuclear charges $z_i$ and distances among all other atoms:
\begin{equation}
    h_i^0 = {\rm Embed}(z_i) + \mathbf{W}_{R}\sum_{j}({\rm RBF}(z_i,z_j,\|\overrightarrow{r_{ij}}\|)) + \mathbf{b}_{R}
\end{equation}
where $\mathbf{W}_{R}$ and $\mathbf{b}_{R}$ are parameters of the linear layer which aligns the dimension of embedding and $\rm RBF$.

\paragraph{Geometry-aware Transformer Layer} We concatenate $\{h_i^0\}_{i=1}^N$ and $\{h_{ij}^0\}$ together as inputs and apply $L$ layer geometry-aware Transformer block to interact nodes and edges.
For $l^{th}$ layer, the inputs are $\{h_i^l\}_{i=1}^N$ and $\{h_{ij}^l\}$.
Transformers in the natural language processing domain \cite{vaswani2017attention,bert} usually encode absolute position information (i.e. word order) by adding positional embedding into the input representations.
For 3D atomic modeling, representations need to be rotation- and translation-invariant or equivariant, so relative position information is favored.
We encode relative position information in each geometry-aware Transformer layer by adding a bias into the self-attention computation:
\begin{equation}
    A_{ij} = \frac{(h_i^l\mathbf{W}_q)(h_j^l\mathbf{W}_k)^T}{\sqrt{d}} + b(i,j)
\end{equation}
\begin{equation}
    a_{ij} = \frac{\exp A_{ij}}{\sum_k\exp A_{ik}}
\end{equation}
\begin{equation}
    h^{l+1}_i = \sum_j (h^l_i\mathbf{W}_v)a_{ij}
\end{equation}
where index $i,j,k$ can represent a node or a edge and $b(i,j)$ is our proposed geometry-aware spatial encoding.
$b(i,j)$ requires to capture the relative geometry information.

Specifically, if $i$ and $j$ are both nodes, the only relative geometry information is their distance (Figure~\ref{fig:head} \textbf{(c)}), so we use RBF of distance to calculate $b(i,j)$:
\begin{equation}
    b(i,j) = {\rm RBF}(z_i,z_j,\|\overrightarrow{r_{ij}}\|)
\end{equation}

An edge $(i,j)$ and a node $k$ form a triangle $i,j,k$, and the relative geometry information is determined by the lengths of two sides $\|\overrightarrow{r_{ik}}\|, \|\overrightarrow{r_{jk}}\|$ and the included angle $\angle ikj$ (Figure~\ref{fig:head} \textbf{(d)}). 
We calculate bias $b(ij,k)$ by RBF of them:
\begin{equation}
    b(ij,k) = {\rm RBF}(\|\overrightarrow{r_{ik}}\|) + {\rm RBF}(\|\overrightarrow{r_{jk}}\|) + {\rm RBF}(\cos \angle ikj)
\end{equation}

Two edges $(i,j)$ and $(k,l)$ connecting different node pairs form a quadrilateral, and the relative geometry information is determined by lengths of six sides (i.e. $(i,j), (i,k), (i,l), (j,k), (j,l),$ and $(k,l)$).
Lengths of $(i,j)$ and $(k,l)$ are already encoded in $h_{ij}^l$ and $h_{kl}^l$. We use the four other sides to encode $b(ij,kl)$. Furthermore, we enhance the spatial encoding by the angle of edges $(i,j)$ and $(k,l)$ to explicitly embed the angle information (Figure~\ref{fig:head} \textbf{(e)}).
\begin{align}
\begin{split}
    b(ij,kl) =& {\rm RBF}(\|\overrightarrow{r_{ik}}\|) + {\rm RBF}(\|\overrightarrow{r_{jk}}\|) +
    {\rm RBF}(\|\overrightarrow{r_{il}}\|) \\ +& {\rm RBF}(\|\overrightarrow{r_{jl}}\|) + {\rm RBF}(\cos(\overrightarrow{r_{ij}},\overrightarrow{r_{kl}}))
\end{split}
\end{align}

The final representations of nodes $h_i$ and edges $h_{ij}$ are obtained by repeatedly sending the representations into a stack of Transformer layers following the work in \cite{alphafold,graphormer3d}.
This procedure improves the model performance without enlarging parameter counts.

\begin{figure*}[ht]
    \centering
    \includegraphics[width=0.8\linewidth]{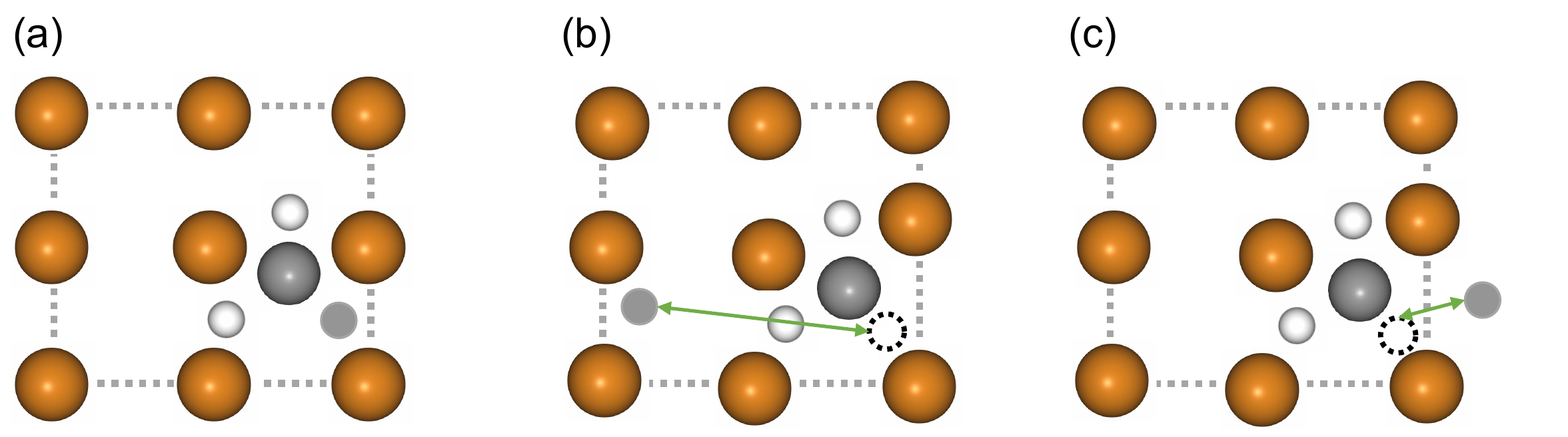}
    \caption{\textbf{PBC Correction.} \textbf{(a)-(c)} An example of a catalyst system where \ch{CH3} is the adsorbate and the surface is composed of copper atoms viewed along the \textless 110\textgreater orientation. The dotted lines show the borders of the unit cell. Grey-colored small atom is a hydrogen atom that goes across the unit cell and the dotted circle is its initial position.
    \textbf{(a)} Initial positions.
    \textbf{(b)} Relaxed positions. The grey hydrogen atom moves across the boundary and reappears on the opposite side.
    \textbf{(c)} PBC correction positions. The grey hydrogen atom lies outside the unit cell and has a smaller distance from the initial position.
    }
    \label{fig:pbc}
\end{figure*}

\paragraph{Predictions}
Inspired by the force field which includes bonds as factors of molecular system energy, we use not only node representations $\{h_i\}$ but also edge representations $\{h_{ij}\}$ for energy (and other scalar properties) predictions:
\begin{equation}
    \hat{y} = \sum_i {\rm MLP}(h_i) + \sum_{ij} {\rm MLP}(h_{ij})
\end{equation}
where $\rm MLP$ is a 2-layer multi-layer perceptron with non-linearity.









\subsection{Periodic boundary condition correction}
\label{pbcc}
Periodic boundary conditions (PBC) are commonly used in molecule system simulation for a large system \cite{pbc}.
To implement PBC, a unit cell (which is also referred as a periodic box) is surrounded by translated copies in all directions to approximate an infinitely large system. 
When one molecule diffuses across the boundary of the periodic box, it reappears on the opposite side. 
Falling to handle such cases will cause errors in atomic distance calculation.
Since distance is one of the most important input features for machine learning methods, this error is nonnegligible.
For example, when considering the relative distance between an initial position $x_i$ and a relaxed position $x_i^r$ in a relaxation process, a long distance will be mistakenly calculated if the relaxed position appears on the opposite side of the boundary. 
To correctly model the local geometry, the relaxed position $x_i^r$ is estimated by finding the smallest possible movement distance from the initial position $x_i$.

\begin{equation}
    \tilde{x}_{i}^r = \mathop{\arg\min}_{x_{i}^r + cell * \delta} (\|x_{i}^r + cell * \delta - x_i\|)
\end{equation}
where $x_i$ is the initial position, $cell$ are three primitive translation vectors defining the unit cell, and $\delta=(\delta_1, \delta_2, \delta_3) \in \mathbf{R}^3$ indicates the directions of atomic movement across boundaries in which directions ($\delta_i\in \{-1,0,1\}$).
We correct the relaxed positions of atoms which do not guarantee relaxed positions are within the unit cell but preserve the correct relative position information between initial positions and relaxed positions.
Figure~\ref{fig:pbc} shows examples of PBC corrections of a catalyst system. In the ablation study, we will show PBC corrections improve the performances on catalyst system energy estimation.





\section{Experiments}

In this section, we show the results of Moleformer tested on the OC20 IS2RE and the QM9 datasets.
Moleformer predicts molecular properties and relaxed energies accurately and outperforms prior arts.

\subsection{Benchmarks on OC20 relaxed energy prediction}
Overall, the OC20 dataset \cite{oc20} is targeted to model the computational catalyst discovery process which is important in solar fuels synthesis, renewable fertilizer production, and energy storage \cite{norskov2013catalyst,norskov2014fundamental,doi:10.1126/science.aad4998}.
OC20 contains large amounts of samples in diverse adsorbates and surfaces.
A large amount of training samples is suitable to test the generation ability of a machine-learning method.

\begin{table*}[!htb]
\small
\caption{Results on OC20 IS2RE \textbf{test} set using direct approach-based methods. The unit for energy MAE is meV, and the unit for Energy Within Threshold (EWT) is percentage. * means ensembled results. ID and OOD mean in-domain and out-of-domain data split. ADS, CAT, and BOTH mean out-of-domain adsorbate, catalyst, and both (i.e. adsorbate and catalyst) data splits respectively. AVG means average results among different splits. Bolded results are the best results among non-ensembled results.} 
\label{table:oc20main}
\begin{center}
\begin{tabular}{lcccccccccc}
\toprule
Methods&\multicolumn{5}{c}{Energy MAE ↓}  &\multicolumn{5}{c}{EWT ↑} \\ 
&ID&\multicolumn{3}{c}{OOD}&AVG&ID&\multicolumn{3}{c}{OOD}&AVG\\
&&ADS&CAT&BOTH&&&ADS&CAT&BOTH&\\
\midrule 
GNS \cite{gnsnn}&422&568&437&465&473&\textbf{9.12}&4.25&\textbf{8.01}&4.64&\textbf{6.51}\\
Graphormer * \cite{graphormer3d}&398&572&417&503&472&8.97&3.45&8.18&3.79&6.10      \\
Equiformer \cite{Equiformer} &417&548&\textbf{425}&474&466&7.71&3.70&7.15&4.07&5.66 \\
\textbf{Moleformer} & \textbf{413} & \textbf{535} & 428 & \textbf{458} & \textbf{459} & 8.79 & \textbf{4.67} & 7.58 & \textbf{4.87} & 6.48  \\
\bottomrule
\end{tabular}
\end{center}
\end{table*}

Specifically, OC20 provides a small molecule called adsorbate composed of H, C, N, and O atoms, and a large surface slab composed of multiple atoms in each sample.
The slab is periodic in all directions with a vacuum layer towards to adsorbate.
OC20 defines three different prediction tasks: S2EF (State to energy and force), IS2RS (Initial state to relaxed state), and IS2RE (Initial state to relaxed energy).
We focus our experiments on IS2RE direct setting which is the most common task in catalysis since the relaxed energy is related to catalyst properties \cite{oc20}.
The direct setting is an end-to-end task that takes the initial state (i.e. atom positions) as inputs and predict the relaxed energy of systems. Relaxed energy is computed via DFT as the targets. 
The models need to compute the relaxed energy from the initial state directly without relaxation. 
Energy mean absolute error (MAE) and energy within threshold (EWT) are used as evaluation metrics. 
The training set contains 470K samples with 46 different adsorbates.
The development set and the test sets are divided into four sub-splits including one in-domain (ID, sampled from the training distribution) split and three out-of-domain (OOD) splits, namely OOD adsorbates (unseen adsorbates), OOD catalysts (unseen element compositions for catalysts), and OOD both (both unseen adsorbate and unseen catalyst compositions).
OOD adsorbates split and OOD both split include 6 different adsorbates: \ch{CHO}, \ch{CH}, \ch{NH2}, \ch{C2H3O2}, \ch{HNO2}, and \ch{C2H4(NH2)2}.
Each sub-split contains 25K samples.
The three OOD splits serve as a good validation of the transportability and generation ability of our proposed architecture.


Notably, the training samples of OC20 are generated by the ASE tool (Atomic Simulation Environment) \cite{ase} which uses PBC to provide atomic positions within the unit cell. 
As mentioned above, atomic positions at the opposite side of the unit cell need to be corrected first for an accurate geometric calculation. Therefore, the proposed PBC correction method is implemented here for data pre-processing, which has been neglected in the previous study \cite{graphormer3d}. Besides, we augment the training data with noisy node data augmentation following \cite{gnsnn,Equiformer}, which first applies a linear interpolation between initial and relaxed positions and adds Gaussian noise on the interpolated positions as input. 
In addition to the main object of relaxed energy prediction in IS2RE, the IS2RS is used as an auxiliary node-level object to predict relaxed states (i.e., structures) \cite{graphormer3d,Equiformer}. Moreover, bond lengths are quadratic terms in the force field of molecular system energy. Given geometric information related to edges is used explicitly in Moleformer, we also add an edge-level object as an auxiliary loss, which needs to predict relaxed edge lengths. Contributions of PBC correction, data augmentation, and especially the auxiliary task of relaxed edge length prediction will be analyzed in the ablation study below. 

We compare Moleformer with GNS \citep{pmlr-v119-sanchez-gonzalez20a,gnsnn}, Graphormer \citep{graphormer3d}, and Equiformer \citep{Equiformer} on IS2RE direct setting.
GNS is a GNN designed for simulating complex physics with an encoder-decoder architecture.
Graphormer modifies Transformer with centrality encoding, spatial encoding, and edge encoding to adapt to graph data.
Equiformer is a rotational- and translational-equivariant Transformer based on irreducible representations.

Results of the test set of OC20 IS2RE are listed in Table~\ref{table:oc20main} and results of the development set are listed in Appendix A.
Moleformer obtains a new direct approach IS2RE state-of-the-art by achieving average energy MAE with 459 meV in the test set, which outperforms GNS with 473 meV, Graphormer with 472 meV, and Equiformer with 466 meV (with an $1.5\%$ relative improvement).
To be noticed, Moleformer outperforms the ensembled models Graphormer with a single model.
For average EWT, Moleformer performs better than Graphormer and Equiformer and performs on par with GNS.
For out-of-domain distribution, Moleformer achieves the lowest energy MAE and the highest EWT in OOD-ADS, and OOD-Both subsets among all methods, which shows Moleformer can better generalize to out-of-domain adsorbates.

\subsection{Benchmarks on QM9 properties prediction} 
The QM9 dataset \cite{qm9} consists of organic molecules made up of H, C, N, O, and F with 3D atom positions. The target of the QM9 dataset is to predict the energetic, electronic, and thermodynamic-related properties.
Different from previous works that consider each target as single-task learning, we experiment Moleformer with a multi-task learning setting which reduces training time greatly.
\citet{gasteiger_dimenet_2020} find that multi-task learning harm the performances on QM9 significantly which does not happen to Moleformer.
Results compared to previous works are listed in Table~\ref{tab:qm9}.
Some works \citep{schnet,gasteiger_dimenet_2020,gasteiger_dimenetpp_2020,painn,torchmd} use random splits with 110K, 10K, and 10K samples in train, development, and test sets respectively. We list their results as references.

\begin{table*}[!htb]
\small
\caption{Results on QM9 \textbf{test} set compared to previous works. Both random split and fix split results are reported.}
\begin{center}
\begin{tabular}{lcccccc}
\toprule
Model & $\alpha$ & $\Delta\epsilon$ & $\rm\epsilon_{HOMO}$ & $\rm\epsilon_{LUMO}$ & $\mu$ & $C_v$  \\
Unit & $a^3_0$ & meV & meV & meV & D & cal/mol K \\
\midrule
\textbf{Random Split} \\
SchNet \cite{schnet}  & .235 & 63 & 41 & 34 & .033 & .033 \\
DimeNet \cite{gasteiger_dimenet_2020} & .047 & 35 & 28 & 20 & .029 & .025 \\
DimeNet++ \cite{gasteiger_dimenetpp_2020} & .044 & 33 & 25 & 20 & .030 & .023 \\
PaiNN \cite{painn} & .045 & 46 &28&20& .012& .024 \\
ET \cite{torchmd} & .059&36& 20&18&.011&.026\\
EQGAT \cite{eqgat} & .053 & 32 & 20 & 16 & .011 & .024 \\
\midrule
\textbf{Fix Split} \\
Cormorant \cite{Cormorant}  & .085 & 61 & 34& 38 &.038& .026 \\
SE(3)-Transformer \cite{se3}  &.142& 53 &35& 33& .051 &.054 \\
EGNN \cite{egnn}  & .071 & 48 & 29 & 25 & .029 & .031 \\
SEGNN \cite{segnn}  & .060& 42 &24 & \textbf{21} & .023 & .031 \\
EQGAT \cite{eqgat}  & .063& 44& 26&22& \textbf{.014}& .027\\
\textbf{Moleformer} & \textbf{.058} & \textbf{30} & \textbf{21} & \textbf{21} & .039 & \textbf{.026} \\

\bottomrule
\end{tabular}
\end{center}
\label{tab:qm9}
\end{table*}

Moleformer outperforms recent state-of-the-art methods including SEGNN \cite{segnn} and EQGAT \cite{eqgat} on five targets.
Moleformer achieves MAE of 0.058 $a^3_0$, 30 meV, 21 meV, 21 meV, 0.039 D, and 0.026 cal/mol K on isotropic polarizability $\alpha$, the gap $\Delta \epsilon$ between $\rm\epsilon_{HOMO}$ and $\rm\epsilon_{LUMO}$, highest occupied molecular orbital energy $\rm\epsilon_{HOMO}$, lowest unoccupied molecular orbital energy $\rm\epsilon_{LUMO}$, dipole moment $\mu$, and heat capacity at 298.15K $C_v$ respectively.
Moleformer has significant improvement on predicting $\Delta \epsilon$, which is MAE of 30 meV compared to MAE of 42 meV from SEGNN.
Moleformer does not achieve state-of-the-art on predicting $\mu$.
\citet{painn} proposes a specific decoder for predicting $\mu$, while we only use a 2-layer MLP. The specific decoder may further enhance the performance of Moleformer.
For more results on QM9 of Moleformer, we refer readers to Appendix B.

\begin{figure*}[!htb]
    \centering
    \includegraphics[width=0.85\linewidth]{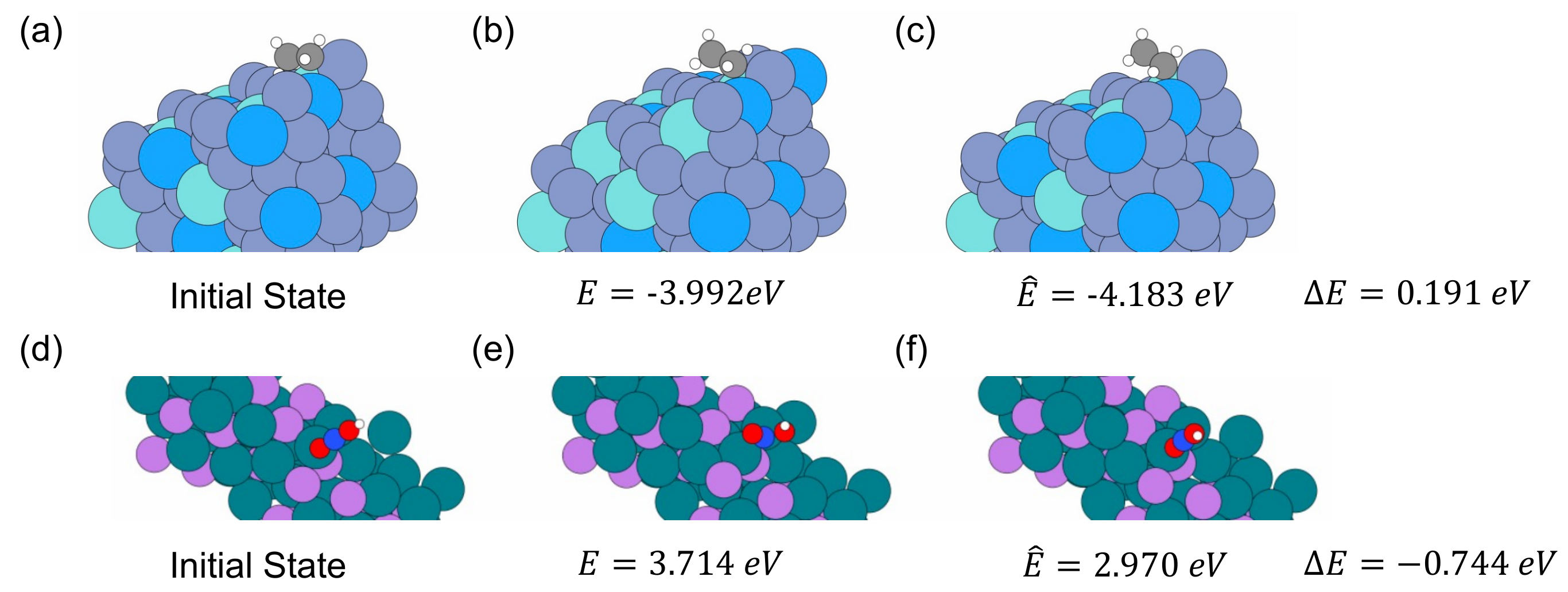}
    \caption{
    \textbf{IS2RS examples for Moleformer.} \textbf{(a),(d)} Initial states. \textbf{(b),(e)} Relaxed states and adsorption energies in relaxed states (i.e. targets). \textbf{(c), (f)} Predicted states and predicted energies from Moleformer. \textbf{(a)-(c)} An in-domain catalyst system with adsorbate \ch{C2H4} and slab composed of Chromium, Zirconium, and Tantalum. \textbf{(d)-(f)} An out-of-domain catalyst system with adsorbate \ch{HNO2} and slab composed of Arsenic and Rhodium.
    }
    \label{fig:is2rs}
\end{figure*}

\section{Discussion}

\subsection{Initial Structure to Relaxed Structure}
To show the performance of molecular dynamic simulations, we visualize Moleformer predicted relaxed structures of catalyst systems from the OC20 dataset based on initial structures in Figure~\ref{fig:is2rs}.
Typically, relaxed structures are simulated by iteratively estimating atomic forces via DFT and updating atom positions.
This may require hundred times for convergence, however, predicting relaxed structures based on initial structures only needs one step inference.
From Figure~\ref{fig:is2rs} \textbf{(a)-(c)}, we find that Moleformer predicts relaxed structures of in-domain adsorbates accurately. The predicted states of \ch{C2H4} are similar to the target relaxed states, and the predicted energy is also close to the target relaxed energy.
For out-of-domain adsorbates shown in Figure~\ref{fig:is2rs} \textbf{(d)-(f)}, the prediction states of Moleformer are not close enough to the relaxed states and the energy difference is higher than in-domain energy estimation.
We can see that Moleformer has the ability for \textit{ab initio} relaxed state prediction, and the task of energy estimation is highly correlated to the task of state estimation.

\begin{figure*}[ht]
    \centering
    \includegraphics[width=0.95\linewidth]{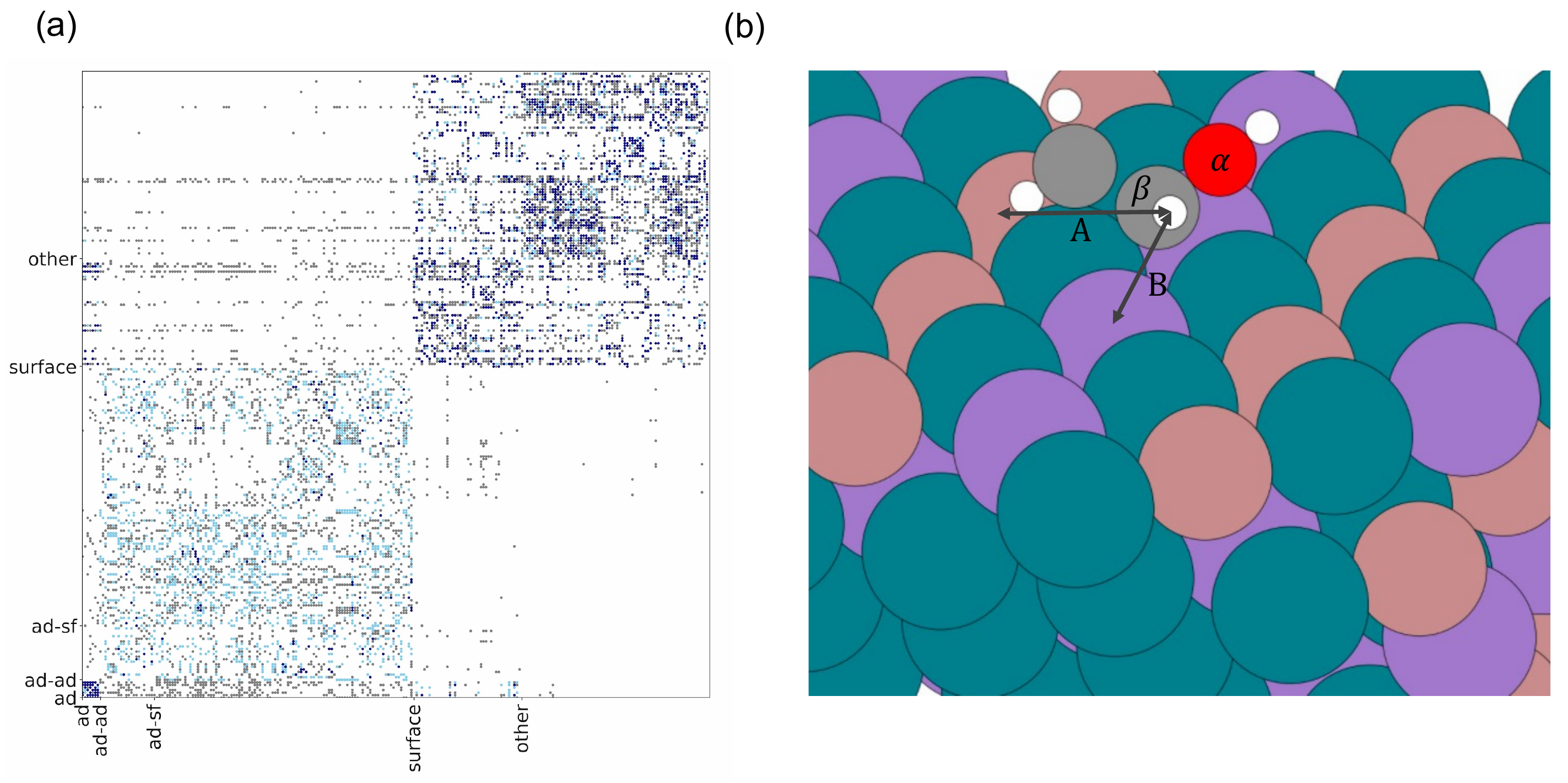}
    \caption{
    A catalyst system sampled from the OC20 dataset where \ch{H2C=CHOH} is the adsorbate molecule (Hydrogen atom in white, Carbon atom in grey, and Oxygen atom in red) and the slab is composed of Manganese (in purple), Rhodium (in dark green), and Gallium (in dark red) atoms. 
    \textbf{(a)} The attention map calculated by Moleformer in this catalyst system. In X and Y-axes, ad means adsorbate and sf means the surface of the slab. The deeper color indicates more attention between these inputs.
    \textbf{(b)} 3D visualization of the catalyst system, two adsorbate-surface edges (A and B) have the highest attention value with an Oxygen atom $\alpha$ and a Carbon atom $\beta$ of the adsorbate.
    }
    \label{fig:yugang}
\end{figure*}

\subsection{Moleformer Interpretation}
To understand how Moleformer models nodes and edges together, we plot the attention map for a sampled catalyst system from OC20 in Figure~\ref{fig:yugang}.
For each input pair, we calculate the attention similarities $A$ between them.
We summarize the attention similarities in each layer and each head by counting the times that attention similarities exceed a threshold.
The deep blue color indicates two inputs interact frequently, the blue color indicates two inputs interact occasionally, and the grey color indicates two inputs interact infrequently.
It can be seen from Figure~\ref{fig:yugang} \textbf{(a)} that adsorbate atoms interact with adsorbate atoms frequently, and slab atoms interact with slab atoms frequently. Adsorbate atoms interact less with slab atoms but interact more with adsorbate-related edges.
This shows Moleformer uses edges to communicate the adsorbate and the slab.
In Figure~\ref{fig:yugang} \textbf{(b)}, we calculate the attention similarities of the last layer from Moleformer. 
The Oxygen atom $\alpha$ has the highest attention similarity to the edge A connecting the Hydrogen atom and the nearest Gallium atom.
The Carbon atom $\beta$ has the highest attention similarity to the edge B connecting the Hydrogen atom and the nearest Manganese atom.
These two edges are nonbonding atom pairs which may be neglected by other machine-learning models which only consider bonded atom pairs.
These attentions involve three different atoms at once, which shows that Moleformer has the ability to model complex atom interactions.

\begin{figure*}[!htb]
    \centering
    \includegraphics[width=0.85\linewidth]{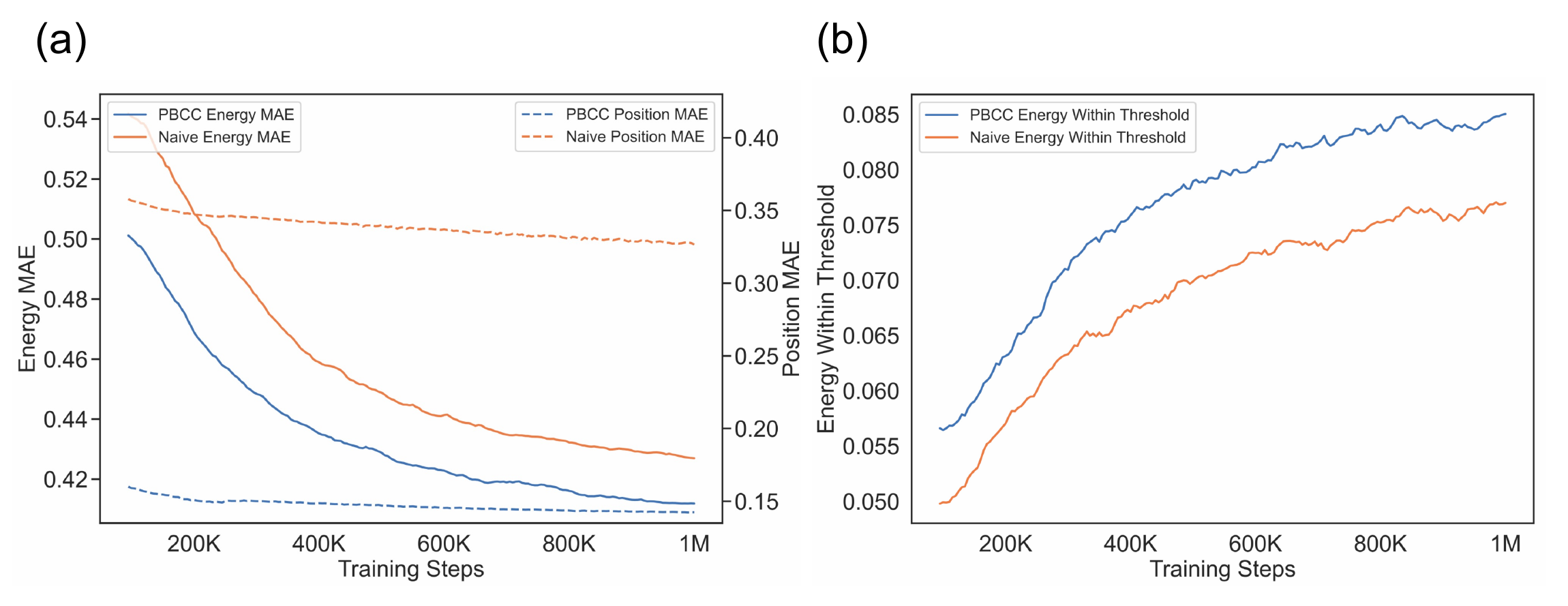}
    \caption{\textbf{Ablation study on PBC Correction.} With PBC correction, Moleformer achieves better position MAE in IS2RS and better energy MAE and EWT in IS2RE. Results of MAE and EWT are obtained in the OC20 in-domain validation set.
    }
    \label{fig:pbcab}
\end{figure*}


\subsection{Ablation Study}
In order to measure the contributions of PBC correction, noisy node data augmentation and an auxiliary edge-level loss, we perform ablation studies on the OC20 dataset. The experiments are conducted repeatedly, with one module removed each time. 
The results are listed in Table~\ref{tab:Ablation}.

As shown in Figure~\ref{fig:pbcab}, PBC corrections reduce the position MAE of IS2RS by calculating moving distances correctly.
As a result, in in-domain split, an energy MAE improvement of 0.014 eV and an EWT improvement of $0.7\%$ in IS2RE are obtained, validating the effectiveness of PBC corrections.
From Table~\ref{tab:Ablation}, we observe training without PBC correction, the energy MAE of the in-domain split and out-of-domain splits increase consistently.

Moreover, without noisy node augmentation, the energy MAE of the out-of-domain adsorbate split increases significantly, while the energy MAE of the out-of-domain catalyst split has a minor change.
The OC20 dataset has only 82 adsorbates (this number is different from the IS2RE dataset since other adsorbates appear in the S2EF dataset) and up to $10^5$ various catalyst slabs.
Training on OC20 is easy to overfit on limited adsorbates.
Noisy node augmentation provides different initial states of adsorbates and improves the generalization abilities.

Given that Moleformer introduces graph edges as inputs, and we add an auxiliary edge-level loss by predicting relaxed edge lengths.
Removing edge-level auxiliary loss increases energy MAE in both in-domain split and out-of-domain splits.
Previous works \cite{gnsnn,graphormer3d,Equiformer} have proven that adding node-level auxiliary loss helps energy estimation, here we find that adding edge-level auxiliary loss also works.
Consider all atoms in a system as a set, the system energy is contributed by all elements in the power set (including atoms, bonds (2 atoms), angles (3 atoms), etc.). 
Since bond lengths are highly related to energy, introducing supervision for bonds naturally helps the estimation of energy.

Overall, adding the auxiliary edge-level loss achieved the largest performance gain among the three modules indicating the effectiveness of integrating geometric information related to edges into the Moleformer architecture.



\begin{table}[!tp]
\small
\centering
\caption{Ablation studies on OC20 IS2RE \textbf{development} set with energy MAE. The unit for energy MAE is meV.}
\begin{tabular}{lccccc}
\toprule
Splits & ID & \multicolumn{3}{c}{OOD} & AVG \\
Settings & & ADS & CAT & BOTH & \\
\midrule
Moleformer& \textbf{413} & \textbf{523} & 432&\textbf{473}&\textbf{460} \\
- PBC Correction &427 & 551 & 441 &495&479 \\
- Data Augmentation&417 & 587 & \textbf{429} & 530 & 491 \\
- Edge Auxiliary Loss &433 & 576 & 441 & 529&495 \\
\bottomrule
\end{tabular}
\label{tab:Ablation}
\end{table}

\section{Conclusion}
This work introduces Moleformer, a novel Transformer architecture to estimate 3D atomic system energy and properties.
Moleformer is inspired by the force field for molecular system energy calculation which considers bonds, bond angles, torsion angles, and nonbonding interactions as factors.
Moleformer leverages atoms and atom pairs as inputs, and models factors above by geometry-aware spatial encoding.
We have shown that Moleformer is more accurate in predicting atomic system energy and properties than other GNN and Transformer models.
GNNs have dominated the 3D atomic system modeling due to the nature of molecules which are composed of atoms (nodes) and bonds (edges).
Transformers are powerful in sequential data, especially in texts, which need specific domain inductive bias to adapt to other data formats.
The inductive biases in Moleformer are (1) interactions among nodes, bonds, and nonbonding atoms contribute to system energy; (2) interactions among several atoms are determined by their atomic numbers, relative distances, and angles.
These inductive biases make Moleformer suited for 3D atomic system modeling.
A possible direction for future work is to introduce higher-order factors that contributed to system energy into the model. 

Moleformer improves molecular energy estimation performance by involving edges and performing the attention mechanism among nodes and edges.
However, this leads to more memory and computational requirements.
Compared to GNNs, Transformer-based models are memory and computationally intensive.
The further development direction of Moleformer can be adapting to a light-weight Transformer.

For machine learning methods, generalization ability is an important factor.
Moleformer shows consistent improvement in in-domain molecular and out-of-domain molecular for energy prediction in the OC20 dataset. 
Moleformer improves out-of-domain molecular by extending edge features and interactions.
However, energy predictions of out-of-domain molecular still have a performance gap with in-domain molecular.
Another direction is to further improve the generalization ability for unseen molecules which will be useful in the design of \textit{de novo} molecular.


\bibliography{example_paper}

\begin{thebibliography}{45}
\providecommand{\natexlab}[1]{#1}
\providecommand{\url}[1]{\texttt{#1}}
\expandafter\ifx\csname urlstyle\endcsname\relax
  \providecommand{\doi}[1]{doi: #1}\else
  \providecommand{\doi}{doi: \begingroup \urlstyle{rm}\Url}\fi

\bibitem[Anderson et~al.(2019)Anderson, Hy, and Kondor]{Cormorant}
Anderson, B., Hy, T.~S., and Kondor, R.
\newblock Cormorant: Covariant molecular neural networks.
\newblock In Wallach, H., Larochelle, H., Beygelzimer, A., d\textquotesingle
  Alch\'{e}-Buc, F., Fox, E., and Garnett, R. (eds.), \emph{Advances in Neural
  Information Processing Systems}, volume~32. Curran Associates, Inc., 2019.
\newblock URL
  \url{https://proceedings.neurips.cc/paper/2019/file/03573b32b2746e6e8ca98b9123f2249b-Paper.pdf}.

\bibitem[Brandstetter et~al.(2021)Brandstetter, Hesselink, van~der Pol,
  Bekkers, and Welling]{segnn}
Brandstetter, J., Hesselink, R., van~der Pol, E., Bekkers, E., and Welling, M.
\newblock Geometric and physical quantities improve e(3) equivariant message
  passing.
\newblock 2021.

\bibitem[Chanussot et~al.(2021)Chanussot, Das, Goyal, Lavril, Shuaibi, Riviere,
  Tran, Heras-Domingo, Ho, Hu, Palizhati, Sriram, Wood, Yoon, Parikh, Zitnick,
  and Ulissi]{oc20}
Chanussot, L., Das, A., Goyal, S., Lavril, T., Shuaibi, M., Riviere, M., Tran,
  K., Heras-Domingo, J., Ho, C., Hu, W., Palizhati, A., Sriram, A., Wood, B.,
  Yoon, J., Parikh, D., Zitnick, C.~L., and Ulissi, Z.
\newblock Open catalyst 2020 (oc20) dataset and community challenges.
\newblock \emph{ACS Catalysis}, 11\penalty0 (10):\penalty0 6059--6072, 2021.
\newblock \doi{10.1021/acscatal.0c04525}.
\newblock URL \url{https://doi.org/10.1021/acscatal.0c04525}.

\bibitem[Cohen et~al.(2012)Cohen, Mori-S{\'a}nchez, and Yang]{dft}
Cohen, A.~J., Mori-S{\'a}nchez, P., and Yang, W.
\newblock Challenges for density functional theory.
\newblock \emph{Chemical reviews}, 112\penalty0 (1):\penalty0 289--320, 2012.

\bibitem[Devlin et~al.(2019)Devlin, Chang, Lee, and Toutanova]{bert}
Devlin, J., Chang, M.-W., Lee, K., and Toutanova, K.
\newblock {BERT}: Pre-training of deep bidirectional transformers for language
  understanding.
\newblock In \emph{Proceedings of the 2019 Conference of the North {A}merican
  Chapter of the Association for Computational Linguistics: Human Language
  Technologies, Volume 1 (Long and Short Papers)}, pp.\  4171--4186,
  Minneapolis, Minnesota, June 2019. Association for Computational Linguistics.
\newblock \doi{10.18653/v1/N19-1423}.
\newblock URL \url{https://aclanthology.org/N19-1423}.

\bibitem[Durrant \& McCammon(2011)Durrant and McCammon]{durrant2011molecular}
Durrant, J.~D. and McCammon, J.~A.
\newblock Molecular dynamics simulations and drug discovery.
\newblock \emph{BMC biology}, 9\penalty0 (1):\penalty0 1--9, 2011.

\bibitem[Frenkel \& Smit(2002)Frenkel and Smit]{md}
Frenkel, D. and Smit, B.
\newblock \emph{Understanding Molecular Simulation: From Algorithms to
  Applications}, volume~1 of \emph{Computational Science Series}.
\newblock Academic Press, San Diego, second edition, 2002.

\bibitem[Fuchs et~al.(2020)Fuchs, Worrall, Fischer, and Welling]{se3}
Fuchs, F., Worrall, D., Fischer, V., and Welling, M.
\newblock Se(3)-transformers: 3d roto-translation equivariant attention
  networks.
\newblock In Larochelle, H., Ranzato, M., Hadsell, R., Balcan, M., and Lin, H.
  (eds.), \emph{Advances in Neural Information Processing Systems}, volume~33,
  pp.\  1970--1981. Curran Associates, Inc., 2020.
\newblock URL
  \url{https://proceedings.neurips.cc/paper/2020/file/15231a7ce4ba789d13b722cc5c955834-Paper.pdf}.

\bibitem[Gasteiger et~al.(2020{\natexlab{a}})Gasteiger, Giri, Margraf, and
  G{\"u}nnemann]{gasteiger_dimenetpp_2020}
Gasteiger, J., Giri, S., Margraf, J.~T., and G{\"u}nnemann, S.
\newblock Fast and uncertainty-aware directional message passing for
  non-equilibrium molecules.
\newblock In \emph{Machine Learning for Molecules Workshop, NeurIPS},
  2020{\natexlab{a}}.

\bibitem[Gasteiger et~al.(2020{\natexlab{b}})Gasteiger, Gro{\ss}, and
  G{\"u}nnemann]{gasteiger_dimenet_2020}
Gasteiger, J., Gro{\ss}, J., and G{\"u}nnemann, S.
\newblock Directional message passing for molecular graphs.
\newblock In \emph{International Conference on Learning Representations
  (ICLR)}, 2020{\natexlab{b}}.

\bibitem[Gasteiger et~al.(2021)Gasteiger, Becker, and G{\"u}nnemann]{gemnet}
Gasteiger, J., Becker, F., and G{\"u}nnemann, S.
\newblock Gemnet: Universal directional graph neural networks for molecules.
\newblock \emph{Advances in Neural Information Processing Systems},
  34:\penalty0 6790--6802, 2021.

\bibitem[Godwin et~al.(2021)Godwin, Schaarschmidt, Gaunt, Sanchez-Gonzalez,
  Rubanova, Velivckovi'c, Kirkpatrick, and Battaglia]{gnsnn}
Godwin, J., Schaarschmidt, M., Gaunt, A., Sanchez-Gonzalez, A., Rubanova, Y.,
  Velivckovi'c, P., Kirkpatrick, J., and Battaglia, P.~W.
\newblock Simple gnn regularisation for 3d molecular property prediction and
  beyond.
\newblock In \emph{International Conference on Learning Representations}, 2021.

\bibitem[Hu et~al.(2021)Hu, Shuaibi, Das, Goyal, Sriram, Leskovec, Parikh, and
  Zitnick]{hu2021forcenet}
Hu, W., Shuaibi, M., Das, A., Goyal, S., Sriram, A., Leskovec, J., Parikh, D.,
  and Zitnick, C.~L.
\newblock Forcenet: A graph neural network for large-scale quantum
  calculations.
\newblock \emph{arXiv preprint arXiv:2103.01436}, 2021.

\bibitem[Jones(2015)]{dft2}
Jones, R.~O.
\newblock Density functional theory: Its origins, rise to prominence, and
  future.
\newblock \emph{Reviews of Modern Physics}, 87:\penalty0 897--923, 2015.

\bibitem[Jumper et~al.(2021)Jumper, Evans, Pritzel, Green, Figurnov,
  Ronneberger, Tunyasuvunakool, Bates, Z{\'i}dek, Potapenko, Bridgland, Meyer,
  Kohl, Ballard, Cowie, Romera-Paredes, Nikolov, Jain, Adler, Back, Petersen,
  Reiman, Clancy, Zielinski, Steinegger, Pacholska, Berghammer, Bodenstein,
  Silver, Vinyals, Senior, Kavukcuoglu, Kohli, and Hassabis]{alphafold}
Jumper, J.~M., Evans, R., Pritzel, A., Green, T., Figurnov, M., Ronneberger,
  O., Tunyasuvunakool, K., Bates, R., Z{\'i}dek, A., Potapenko, A., Bridgland,
  A., Meyer, C., Kohl, S. A.~A., Ballard, A., Cowie, A., Romera-Paredes, B.,
  Nikolov, S., Jain, R., Adler, J., Back, T., Petersen, S., Reiman, D.~A.,
  Clancy, E., Zielinski, M., Steinegger, M., Pacholska, M., Berghammer, T.,
  Bodenstein, S., Silver, D., Vinyals, O., Senior, A.~W., Kavukcuoglu, K.,
  Kohli, P., and Hassabis, D.
\newblock Highly accurate protein structure prediction with alphafold.
\newblock \emph{Nature}, 596:\penalty0 583 -- 589, 2021.

\bibitem[Karplus \& Petsko(1990)Karplus and Petsko]{karplus1990molecular}
Karplus, M. and Petsko, G.~A.
\newblock Molecular dynamics simulations in biology.
\newblock \emph{Nature}, 347\penalty0 (6294):\penalty0 631--639, 1990.

\bibitem[Larsen et~al.(2017)Larsen, Mortensen, Blomqvist, Castelli,
  Christensen, Du{\l}ak, Friis, Groves, Hammer, Hargus, et~al.]{ase}
Larsen, A.~H., Mortensen, J.~J., Blomqvist, J., Castelli, I.~E., Christensen,
  R., Du{\l}ak, M., Friis, J., Groves, M.~N., Hammer, B., Hargus, C., et~al.
\newblock The atomic simulation environment—a python library for working with
  atoms.
\newblock \emph{Journal of Physics: Condensed Matter}, 29\penalty0
  (27):\penalty0 273002, 2017.

\bibitem[Le et~al.(2022)Le, No{\'e}, and Clevert]{eqgat}
Le, T., No{\'e}, F., and Clevert, D.-A.
\newblock Equivariant graph attention networks for molecular property
  prediction.
\newblock \emph{arXiv preprint arXiv:2202.09891}, 2022.

\bibitem[Leach(2001)]{cookbook}
Leach, A.
\newblock \emph{Molecular Modelling: Principles and Applications}.
\newblock Prentice Hall, 2001.
\newblock ISBN 9780582382107.
\newblock URL \url{https://books.google.com/books?id=kB7jsbV-uhkC}.

\bibitem[Liao \& Smidt(2022)Liao and Smidt]{Equiformer}
Liao, Y. and Smidt, T.~E.
\newblock Equiformer: Equivariant graph attention transformer for 3d atomistic
  graphs.
\newblock \emph{ArXiv}, abs/2206.11990, 2022.

\bibitem[Liu et~al.(2021)Liu, Lin, Jia, Cheng, Jiang, Guo, and
  Ma]{doi:10.1021/acs.jcim.0c01224}
Liu, Z., Lin, L., Jia, Q., Cheng, Z., Jiang, Y., Guo, Y., and Ma, J.
\newblock Transferable multilevel attention neural network for accurate
  prediction of quantum chemistry properties via multitask learning.
\newblock \emph{Journal of Chemical Information and Modeling}, 61\penalty0
  (3):\penalty0 1066--1082, 2021.
\newblock \doi{10.1021/acs.jcim.0c01224}.
\newblock URL \url{https://doi.org/10.1021/acs.jcim.0c01224}.
\newblock PMID: 33629839.

\bibitem[Loshchilov \& Hutter(2019)Loshchilov and
  Hutter]{loshchilov2017decoupled}
Loshchilov, I. and Hutter, F.
\newblock Decoupled weight decay regularization.
\newblock In \emph{7th International Conference on Learning Representations,
  {ICLR} 2019, New Orleans, LA, USA, May 6-9, 2019}, 2019.

\bibitem[Luding(2005)]{luding2005molecular}
Luding, S.
\newblock Molecular dynamics simulations of granular materials.
\newblock In \emph{The physics of granular media}, pp.\  297--324. Wiley, 2005.

\bibitem[Makov \& Payne(1995)Makov and Payne]{pbc}
Makov, G. and Payne, M.
\newblock Periodic boundary conditions in ab initio calculations.
\newblock \emph{Physical Review B}, 51\penalty0 (7):\penalty0 4014, 1995.

\bibitem[N{\o}rskov \& Bligaard(2013)N{\o}rskov and
  Bligaard]{norskov2013catalyst}
N{\o}rskov, J.~K. and Bligaard, T.
\newblock The catalyst genome, 2013.

\bibitem[N{\o}rskov et~al.(2014)N{\o}rskov, Studt, Abild-Pedersen, and
  Bligaard]{norskov2014fundamental}
N{\o}rskov, J.~K., Studt, F., Abild-Pedersen, F., and Bligaard, T.
\newblock \emph{Fundamental concepts in heterogeneous catalysis}.
\newblock John Wiley \& Sons, 2014.

\bibitem[Paszke et~al.(2019)Paszke, Gross, Massa, Lerer, Bradbury, Chanan,
  Killeen, Lin, Gimelshein, Antiga, Desmaison, Kopf, Yang, DeVito, Raison,
  Tejani, Chilamkurthy, Steiner, Fang, Bai, and Chintala]{pytorch}
Paszke, A., Gross, S., Massa, F., Lerer, A., Bradbury, J., Chanan, G., Killeen,
  T., Lin, Z., Gimelshein, N., Antiga, L., Desmaison, A., Kopf, A., Yang, E.,
  DeVito, Z., Raison, M., Tejani, A., Chilamkurthy, S., Steiner, B., Fang, L.,
  Bai, J., and Chintala, S.
\newblock Pytorch: An imperative style, high-performance deep learning library.
\newblock In \emph{Advances in Neural Information Processing Systems 32}, pp.\
  8024--8035. Curran Associates, Inc., 2019.

\bibitem[Ramakrishnan et~al.(2014)Ramakrishnan, Dral, Rupp, and
  Von~Lilienfeld]{qm9}
Ramakrishnan, R., Dral, P.~O., Rupp, M., and Von~Lilienfeld, O.~A.
\newblock Quantum chemistry structures and properties of 134 kilo molecules.
\newblock \emph{Scientific data}, 1\penalty0 (1):\penalty0 1--7, 2014.

\bibitem[Sanchez-Gonzalez et~al.(2020)Sanchez-Gonzalez, Godwin, Pfaff, Ying,
  Leskovec, and Battaglia]{pmlr-v119-sanchez-gonzalez20a}
Sanchez-Gonzalez, A., Godwin, J., Pfaff, T., Ying, R., Leskovec, J., and
  Battaglia, P.
\newblock Learning to simulate complex physics with graph networks.
\newblock In III, H.~D. and Singh, A. (eds.), \emph{Proceedings of the 37th
  International Conference on Machine Learning}, volume 119 of
  \emph{Proceedings of Machine Learning Research}, pp.\  8459--8468. PMLR,
  13--18 Jul 2020.
\newblock URL
  \url{https://proceedings.mlr.press/v119/sanchez-gonzalez20a.html}.

\bibitem[Satorras et~al.(2021)Satorras, Hoogeboom, and Welling]{egnn}
Satorras, V.~G., Hoogeboom, E., and Welling, M.
\newblock E(n) equivariant graph neural networks.
\newblock In \emph{International Conference on Machine Learning}, 2021.

\bibitem[Sch{\"u}tt et~al.(2021)Sch{\"u}tt, Unke, and Gastegger]{painn}
Sch{\"u}tt, K.~T., Unke, O.~T., and Gastegger, M.
\newblock Equivariant message passing for the prediction of tensorial
  properties and molecular spectra.
\newblock In \emph{International Conference on Machine Learning}, 2021.

\bibitem[Schütt et~al.(2018)Schütt, Sauceda, Kindermans, Tkatchenko, and
  Müller]{schnet}
Schütt, K.~T., Sauceda, H.~E., Kindermans, P.-J., Tkatchenko, A., and Müller,
  K.-R.
\newblock Schnet – a deep learning architecture for molecules and materials.
\newblock \emph{The Journal of Chemical Physics}, 148\penalty0 (24):\penalty0
  241722, 2018.
\newblock \doi{10.1063/1.5019779}.
\newblock URL \url{https://doi.org/10.1063/1.5019779}.

\bibitem[Seh et~al.(2017)Seh, Kibsgaard, Dickens, Chorkendorff, Nørskov, and
  Jaramillo]{doi:10.1126/science.aad4998}
Seh, Z.~W., Kibsgaard, J., Dickens, C.~F., Chorkendorff, I., Nørskov, J.~K.,
  and Jaramillo, T.~F.
\newblock Combining theory and experiment in electrocatalysis: Insights into
  materials design.
\newblock \emph{Science}, 355\penalty0 (6321):\penalty0 eaad4998, 2017.
\newblock \doi{10.1126/science.aad4998}.
\newblock URL \url{https://www.science.org/doi/abs/10.1126/science.aad4998}.

\bibitem[Shazeer(2020)]{glu}
Shazeer, N.
\newblock Glu variants improve transformer.
\newblock \emph{arXiv preprint arXiv:2002.05202}, 2020.

\bibitem[Shi et~al.(2022)Shi, Zheng, Ke, Shen, You, He, Luo, Liu, He, and
  Liu]{graphormer3d}
Shi, Y., Zheng, S., Ke, G., Shen, Y., You, J., He, J., Luo, S., Liu, C., He,
  D., and Liu, T.-Y.
\newblock Benchmarking graphormer on large-scale molecular modeling datasets.
\newblock \emph{arXiv preprint arXiv:2203.04810}, 2022.
\newblock URL \url{https://arxiv.org/abs/2203.04810}.

\bibitem[Shuaibi et~al.(2021)Shuaibi, Kolluru, Das, Grover, Sriram, Ulissi, and
  Zitnick]{spinconv}
Shuaibi, M., Kolluru, A., Das, A., Grover, A., Sriram, A., Ulissi, Z., and
  Zitnick, C.~L.
\newblock Rotation invariant graph neural networks using spin convolutions.
\newblock \emph{arXiv preprint arXiv:2106.09575}, 2021.

\bibitem[Shui \& Karypis(2020)Shui and Karypis]{hmgnn}
Shui, Z. and Karypis, G.
\newblock Heterogeneous molecular graph neural networks for predicting molecule
  properties.
\newblock In \emph{2020 IEEE International Conference on Data Mining (ICDM)},
  pp.\  492--500, Los Alamitos, CA, USA, nov 2020. IEEE Computer Society.
\newblock \doi{10.1109/ICDM50108.2020.00058}.
\newblock URL
  \url{https://doi.ieeecomputersociety.org/10.1109/ICDM50108.2020.00058}.

\bibitem[Th{\"o}lke \& Fabritiis(2022)Th{\"o}lke and Fabritiis]{torchmd}
Th{\"o}lke, P. and Fabritiis, G.~D.
\newblock Equivariant transformers for neural network based molecular
  potentials.
\newblock In \emph{International Conference on Learning Representations}, 2022.
\newblock URL \url{https://openreview.net/forum?id=zNHzqZ9wrRB}.

\bibitem[Van~Gunsteren \& Berendsen(1990)Van~Gunsteren and
  Berendsen]{van1990computer}
Van~Gunsteren, W.~F. and Berendsen, H.~J.
\newblock Computer simulation of molecular dynamics: methodology, applications,
  and perspectives in chemistry.
\newblock \emph{Angewandte Chemie International Edition in English},
  29\penalty0 (9):\penalty0 992--1023, 1990.

\bibitem[Vaswani et~al.(2017)Vaswani, Shazeer, Parmar, Uszkoreit, Jones, Gomez,
  Kaiser, and Polosukhin]{vaswani2017attention}
Vaswani, A., Shazeer, N., Parmar, N., Uszkoreit, J., Jones, L., Gomez, A.~N.,
  Kaiser, {\L}., and Polosukhin, I.
\newblock Attention is all you need.
\newblock \emph{Advances in neural information processing systems}, 30, 2017.

\bibitem[Wang et~al.(2022)Wang, Wang, Zhao, Xu, Hao, Hsieh, Gu, and
  Duan]{wang2022heterogeneous}
Wang, Z., Wang, C., Zhao, S., Xu, Y., Hao, S., Hsieh, C.~Y., Gu, B.-L., and
  Duan, W.
\newblock Heterogeneous relational message passing networks for molecular
  dynamics simulations.
\newblock \emph{npj Computational Materials}, 8\penalty0 (1):\penalty0 1--9,
  2022.

\bibitem[Ying et~al.(2021)Ying, Cai, Luo, Zheng, Ke, He, Shen, and
  Liu]{graphormer}
Ying, C., Cai, T., Luo, S., Zheng, S., Ke, G., He, D., Shen, Y., and Liu, T.-Y.
\newblock Do transformers really perform bad for graph representation?
\newblock In \emph{NeurIPS}, 2021.

\bibitem[Zhang \& Sennrich(2019)Zhang and Sennrich]{rms}
Zhang, B. and Sennrich, R.
\newblock Root mean square layer normalization.
\newblock \emph{Advances in Neural Information Processing Systems}, 32, 2019.

\bibitem[Zhang et~al.(2020)Zhang, Liu, and Xie]{zhang2020molecular}
Zhang, S., Liu, Y., and Xie, L.
\newblock Molecular mechanics-driven graph neural network with multiplex graph
  for molecular structures.
\newblock In \emph{NeurIPS-W}, 2020.

\bibitem[Zitnick et~al.(2022)Zitnick, Das, Kolluru, Lan, Shuaibi, Sriram,
  Ulissi, and Wood]{scn}
Zitnick, C.~L., Das, A., Kolluru, A., Lan, J., Shuaibi, M., Sriram, A., Ulissi,
  Z., and Wood, B.
\newblock Spherical channels for modeling atomic interactions.
\newblock \emph{arXiv preprint arXiv:2206.14331}, 2022.

\end{thebibliography}
\bibliographystyle{icml2023}

\newpage
\appendix
\onecolumn

\section{Training Details}
The proof of rotational and translational invariance for Moleformer is naive since all positional features (i.e. atom pair lengths and angle among vectors) used in Moleformer are rotational and translational invariant.

\paragraph{Architeture Details} We apply the layer normalization and the feed-forward network after the multi-head attention in each Transformer layer.
We adopt root mean square (RMS) layer normalization \cite{rms} following \citet{Equiformer}.
We use GEGLU \cite{glu} as the activation function in the feed-forward network.

\paragraph{OC20} For training on the OC20 IS2RE dataset, we applied noisy nodes data augmentation following \cite{gnsnn} which augment the initial positions by linear interpolation between the initial positions and relaxed positions and add Gaussian noises $N(0,\sigma^2)$ to adsorbate atoms and surface atoms of the slab where $\sigma=0.3 \angstrom$.

For edges used as inputs, we calculated distances between all node pairs.
In the OC20 dataset, adsorbate molecules have few atoms while slabs have more atoms.
If we use all node pairs as inputs, edges will be dominated by two atoms from slabs.
Interactions among slab atoms are monotone, and involving too many edges from two slab atoms is not informative.
We group nodes (i.e. atoms) into three groups: adsorbate atoms, surface atoms of the slab, and other atoms of the slab.
Since we wanted to pay more attention to adsorbates and interactions among adsorbates and surface atoms of slabs, we group all node pairs into six groups based on node types: adsorbate-adsorbate, adsorbate-surface, surface-surface, adsorbate-other, surface-other, and other-other.
We sorted all node pairs based on group types and distances and selected the top $M$ node pairs as edges.
We heuristically selected $M$ equal to the node count, which will ignore some other-other edges which are not informative.
Attention mechanism in Transformers models have $O(N^2)$ time complexity, adding $M=N$ edges into inputs will not change the order of time complexity.

The models were trained with IS2RE loss (system level), IS2RS auxiliary loss (node level), and relaxed edge length prediction auxiliary loss (edge level). These losses are calculated using MAE and weighted summed as the total loss:
\begin{equation}
\begin{split}
    \mathscr{L} = \lambda_S |\hat{E}-E|+  \frac{\lambda_{N}}{3N}\sum_{i=1}^N\sum_{\alpha=1}^3|\hat{x}_{i,\alpha}^r-x^r_{i,\alpha}|
    + \frac{\lambda_{E}}{M}\sum_{e=1}^{M}|\hat{l}_e-l_e|
\end{split}
\end{equation}
where $N$ and $M$ are node count and edge count, $E$ is the relaxed system energy, $x^r_{i,\alpha}$ is the relaxed position of $i^{th}$ atom on direction $\alpha$, $l_e$ is the relaxed length of $e^{th}$ edge, $\hat{\cdot}$ is the prediction from the model, and $\lambda_S$, $\lambda_{N}$, and $\lambda_{E}$ are weights of system, node, and edge level loss respectively.
For training IS2RS auxiliary task, the model needs to predict the moving position of each atom $\Delta \mathbf{x}_i = \mathbf{x}_i^r - \mathbf{x}_i \in \mathbb{R}^3$, where $\mathbf{x}_i^r$ is the relaxed position. 
We apply a rotational equivariant attention layer from \cite{graphormer3d} to predict $\Delta \mathbf{x}_i$ which multiplies unit relative position  $[\frac{\overrightarrow{r_{ij}}}{\|\overrightarrow{r_{ij}}\|}]_{i,j}$ by standard multi-head attention probabilities.
For the relaxed edge length prediction auxiliary task, we use a 2-layer MLP on $h_{ij}$ to regress the relaxed edge length $l$.
The prediction targets of $E$ and $x^r_{i,\alpha}$ had been normalized to zero mean and unit variance.
$\lambda_S$ was set to 1. The weights of auxiliary losses $\lambda_{N}$ and $\lambda_{E}$ were set to 15 and 5 at the beginning of the training and linearly decayed to 1 within the first quarter of training and remained at 1.

\paragraph{QM9}
For the QM9 dataset, we use the split from \cite{Cormorant} which contain 100K, 18K, and 13K samples in train, development, and test sets respectively. 

For edges used as inputs, we select the top $M=N$ nearest node pairs as input edges. 
We apply multi-task training on all 12 targets of QM9, and the total losses are calculated by summing of Smooth L1 Loss for each target.
Smooth L1 Loss performs better than MSE and MAE Loss during our preliminary experiments.
We first apply a linear regression with nuclear charges and targets on the training set, and the coefficients of the linear regression are used as atom references.
For training Moleformer, we subtracted atom references for each target and normalize each target to zero mean and unit variance.
Subtracting atom references significantly reduces the target variances and helps models converge quickly.

\paragraph{Optimization} All models are implemented in PyTorch \cite{pytorch} and trained with 8 NVIDIA A100 GPUs.
We use AdamW \cite{loshchilov2017decoupled} to optimize models with a linear learning rate warmup and decay.
We evaluate the model at the end of training.

\section{Results of development set in the OC20 dataset}
We display the results of the development set of OC20 IS2RE in Table~\ref{table:oc20dev}.
To be noticed, Equiformer with noisy nodes data augmentation seems overfitted on the development set which has a much larger difference between the development set and the test set energy MAE than other methods.

\section{Results of remaining targets in the QM9 dataset}
We display the results of six other targets in Table~\ref{tab:qm9_appendix}.
Moleformer achieves five of six targets' best results among the fix split setting. Moleformer does not perform well in predicting $\langle R^2\rangle$ which may be improved via using a specialized decoder introduced by PaiNN \cite{painn}.

\section{Hyper-parameters for training Moleformer}
We display the used hyper-parameters for training Moleformer in Table~\ref{tab:hyper}.

\begin{table*}[h]
\small
\caption{Results on OC20 IS2RE development set using direct approach-based methods. The unit for energy MAE is meV, and the unit for Energy Within the Threshold (EWT) is percentage. $\dagger$ indicates using noisy nodes data augmentation. ID and OOD mean in-domain and out-of-domain data split. ADS, CAT, and BOTH mean out-of-domain adsorbate, catalyst, and both data splits. AVG means average results among different splits.} 
\label{table:oc20dev}
\begin{center}
\begin{tabular}{lcccccccccc}
\toprule
Methods&\multicolumn{5}{c}{Energy MAE ↓}  &\multicolumn{5}{c}{EWT ↑} \\ 
&ID&\multicolumn{3}{c}{OOD}&AVG&ID&\multicolumn{3}{c}{OOD}&AVG\\
&&ADS&CAT&BOTH&&&ADS&CAT&BOTH&\\
\midrule 
GNS \cite{gnsnn}& 540 & 650 & 550 & 590& 583 & - & - & - & - & - \\
GNS $\dagger$& 470& 510& 480 & 460 &480& - & - & - & - & - \\
Graphormer \cite{graphormer3d}&  433& 585& 444& 530& 498 &   - & - & - & - & -   \\
Equiformer \cite{Equiformer}& 422& 542& 423& 475& 466 &7.23& 3.77& 7.13& 4.10& 5.56\\
Equiformer $\dagger$&416& \textbf{498}& \textbf{417}& \textbf{434}& \textbf{441} &7.47& \textbf{4.64}& 7.19& \textbf{4.84}& \textbf{6.04}\\
\textbf{Moleformer} $\dagger$ & \textbf{413} & 523 & 432&473&460 & \textbf{8.01} & 3.04 & \textbf{7.66} & 3.19 & 5.48 \\
\bottomrule
\end{tabular}
\end{center}
\end{table*}

\begin{table*}[h]
\small
\caption{Results on remaining targets of QM9 test set compared to previous works.}
\begin{center}
\begin{tabular}{lcccccc}
\toprule
Model & $U_0$ & $U$ & $H$ & $G$ & $\langle R^2\rangle$ & ZPVE \\
Unit & meV & meV & meV & meV & $a_0^2$ & meV \\
\midrule
\textbf{Random Split} \\
SchNet \cite{schnet}   & 14 & 19 & 14 & 14 & .073 & 1.7 \\
DimeNet \cite{gasteiger_dimenet_2020}  & 8 & 8 & 8 & 9 & .331 & 1.3 \\
DimeNet++ \cite{gasteiger_dimenetpp_2020}  & 6 & 6 & 7 & 9 & .331 & 1.2 \\
PaiNN \cite{painn}  & 6 & 6 & 6 & 7 & .066 & 1.3\\
ET \cite{torchmd}  & 6 & 6 & 6 & 8 & .033 & 1.8\\
EQGAT \cite{eqgat}  & 25 & 25 & 24 & 23 & .382 & 2.0 \\
\midrule
\textbf{Fix Split} \\
Cormorant \cite{Cormorant} & 22 & 21 & 21 & 20 & .961 & 2.0 \\
EGNN \cite{egnn}  & 12 & 12 & 12 & 12 & .106 & 1.6 \\
SEGNN \cite{segnn} & 15 & 13 & 16 & 15 & .660 & 1.6  \\
EQGAT \cite{eqgat}   & 13 & 13 & 13 & 12 & .257 & 1.5 \\
\textbf{Moleformer} & \textbf{10} & \textbf{10} & \textbf{12} & \textbf{11} & 1.848 & \textbf{1.2} \\
\bottomrule
\end{tabular}
\end{center}
\label{tab:qm9_appendix}
\end{table*}

\begin{table}[h]
\centering
\small
\caption{Hyper-parameters used for training Moleformer on the OC20 dataset and the QM9 dataset.}
\begin{tabular}{lcc}
\toprule
Parameters & OC20 & QM9 \\
\midrule
Embedding Dimension & 768 & 768 \\
Feed-forward Network Dimension & 768 & 768 \\
RBF Dimension & 128 & 128 \\
Attention Heads & 48 & 48 \\
Transformer Layers & 12 & 6  \\
Repeat Count & 4 & 4 \\
Epochs & 70 & 800  \\
Warmup Steps & 10,000 & 1,000 \\
Peak Learning Rate & 2e-4 & 5e-4 \\
Batch Size & 32 & 256  \\
Adam $\epsilon$ & 1e-6 & 1e-6  \\
Weight Decay & 1e-3 & 1e-3 \\
Clipping Gradient & 5 & 5  \\
\bottomrule
\end{tabular}

\label{tab:hyper}
\end{table}

\end{document}